# Trustworthy AI: UK Air Traffic Control Revisited

Rob Procter

Warwick University and Alan Turing Institute for Data Science and AI

rob.procter@warwick.ac.uk

Mark Rouncefield

University of Siegen and Alan Turing Institute for Data Science and AI

mrouncefield@turing.ac.uk

**ABSTRACT**

Exploring the socio-technical challenges confronting the adoption of AI in organisational settings is something that has so far been largely absent from the related literature. In particular, research into requirements for trustworthy AI typically overlooks how people deal with the problems of trust in the tools that they use as part of their everyday work practices. This article presents some findings from an ongoing ethnographic study of how current tools are used in air traffic control work and what it reveals about requirements for trustworthy AI in air traffic control and other safety-critical application domains.

## 1  INTRODUCTION

Project Bluebird[1] is developing agent-based tools with the goal of demonstrating that these AI technologies are capable of achieving progressively higher levels of competence in managing aircraft safely as they pass through controlled air space. The safety critical demands of decision-making in air traffic control make it a particularly perspicuous setting for exploring what trusting technology means as a mundane achievement.

A particular objective for this study is to gain an understanding of role of trust in air traffic control work and how this is manifest in the work practices of air traffic controllers (ATCOs), i.e., trust that individual ATCOs have in one another, trust in the tools they currently use, trust in the wider 'system' of air traffic control, and what this may tell us about trust issues in regard to new AI technologies such as agents.

Understanding trust requires a socio-technical approach, which, for this study means a focus on how the 'lived work' of air traffic controlling is shaped by and, in turn, shapes the socio-material relationships that are embodied in and through ATCOs' everyday work practices and the tools they use.

We begin by briefly reviewing the recent literature on trust in technology. We then draw on some of this study's empirical findings to introduce the concept of 'boundaries', 'contours' or 'gradients' of trust. Our argument is that trust in technologies is not a binary and static relation as it is often presented in the literature but is nuanced and is continuously calibrated and re-calibrated in and through the individual and collective, lived experience of their users.

## 2  TRUST IN TECHNOLOGY

In recent years, the literature of trust in technology and in trust in IT systems, in particular, has begun to engage with "the social mechanisms which generate trust" (Luhmann, 1990: 95). In safety-critical domains this has been driven by an interest on how dependable system performance is achieved:

> "... we are interested in the *in-vivo* work of living with systems that are more or less

---

[1] https://www.nats.aero/about-us/research/n/project-bluebird/

reliable and the practices that this being 'more or less dependable' occasions. The situated practical actions of living with systems (e.g., workarounds and so on) are important… in that they show how those responsible for ensuring dependable operations experience dependability as a practical, day-to-day matter." (Voss et al., 2007: 195)

Trust in technology is a socio-material relation between people, their work practices and the tools they use to assist them that is both 'constitutive' of, and resultant from forms of interaction and organisational work (Watson, 2009). Trust in technology is not an absolute or binary property but has contours that mark out trust gradients defining regions of higher or lower trust. How trust is calibrated, interrogated and its gradients navigated as a "mundane feature of everyday work" (Clarke et al., 2007: 4) is influenced, moment-by-moment, by the circumstances in which users encounter these tools and make sense of their behaviour as a matter of 'ordinary action' (Voss et al., 2007).

Like several previous studies (Kaur et al., 2022; Blazevic et al., 2024), our study is motivated by the prospects of innovative technologies and tools being introduced into established work settings and practices – in this case, agent-based AI systems in air traffic control – and a concern with whether, as technology and work changes, different forms and problems of trust emerge (Thornton et al., 2021). Our interest is in understanding the trust that exists within the current socio-technical system that is air traffic control, how this trust is achieved and sustained, and how the real-time, dependable delivery of air traffic control is reliant on this trust.

Surprisingly, perhaps, given the longstanding interest in the dependability of safety critical systems, there has only been one previous study of UK air traffic control work (Bentley et al., 1992; Hughes et al., 1992). This groundbreaking study, which marked the beginning of the 'turn to the social' in ICT systems requirements gathering (Anderson et al., 1994) drew attention to, *inter alia*, how paper flight strips, which together with radar, were the principal tools in use at that time,[2] afforded dependable air traffic controlling. The ways in which ATCOs collaborate and how these are influenced by interpersonal trust, i.e., ATCOs' trust in one another, were highlighted in this study and in a study of US air traffic control undertaken in the period 1998-2003 (Vaughan, 2021). However, perhaps because of the nature of the technologies then in use, trust in technologies did not emerge as a significant issue in these studies.

The tools used in UK air traffic control have since changed significantly and include those that offer decision support to ATCOs for conflict detection. Our study reveals that interpersonal trust remains a critically important factor for air traffic control work. It also reveals that knowing how ATCOs determine the trustworthiness of the tools they use is essential for understanding how safe and dependable air traffic control work is achieved and will be important for understanding trust requirements for future generations of decision support technologies in air traffic control and other safety critical application domains.

## 3 METHODOLOGY

Ethnography is the study of how people conduct themselves in their natural settings (Hughes, 2001). The aim is to reveal what might otherwise be taken for granted, i.e., the unfolding structure of activities as they are organised by participants, the working divisions of labour, the role of formal and informal collaboration, and how the technologies and tools available are actually used in practice (Garfinkel, 1967; Randall et al., 2021).

As a method, ethnography involves observation and discussions with participants. It has become a commonly used method for gathering requirements for new IT systems since the 'turn to the social' in the early 1990s (Randall et al., 2007). For Project Bluebird, the use of ethnography affords learning about the role of different kinds of trust (i.e., interpersonal trust, trust in technologies, trust in the wider socio-material system that is air traffic control) and how they are calibrated and sustained by the "actors who collectively create the circumstances, contexts, and consequences of technology use." (Dourish, 2006).

---

[2] Paper flight strips survived as a key tool in UK air traffic controlling until 2019.

In this study, observations of ATCOs at work have been conducted in two UK air traffic operations rooms and in sessions where trainee ATCOs undergo instruction in simulated environments. A series of discussions have also been held with 12 ATCOs during operations rooms visits and with instructors at a national air traffic training college. Discussions lasted 60 minutes on average. Notes were taken of what was observed and all discussions were recorded and transcribed before being analysed for key themes.

The objective has been to gain an understanding of what being a competent ATCO means as a practical, everyday achievement, how this is reflected in their work practices, what tools are used and the role that trust plays in how ATCOs use them. Ultimately, the goal is to use what is learnt to contribute to the development of agent-based AI systems that can help to address the future challenges for air traffic control (Blackwell 2021, Marda and Narayan 2021).

## 4 FINDINGS

### 4.1 Collaboration in air traffic control work

When on they are shift in the operations room, ATCOs are organised into 'watches', that is a group of ATCOs, assistants, engineers and other operational support staff, who follow the same repeating shift pattern, starting and finishing that pattern on the same day. In the UK, a watch includes a watch manager; a DOS (Deputy Operational Supervisor); several GSs (Group Supervisors); ATCOs, trainee ATCOs (who are continuing their training in a live environment); around 5 ATSAs (Air Traffic Services Assistants). A specific watch has responsibility for a group of sectors, which are organised by geography and sector type, with each one being the responsibility of an individual ATCO.

An aircraft may pass through one or more adjoining sectors on its route to its destination. Passing from one sector to another involves a transfer of control from the ATCO controlling the sector the aircraft is leaving to the ATCO controlling the sector that will be receiving the aircraft. This is known as a 'coordination' and requires that the two ATCOs to agree on the heading and flight level that the aircraft will enter the receiving sector. Failure of an ATCO to handover the aircraft to the receiving sector as it was coordinated is considered to be 'bad' controlling.

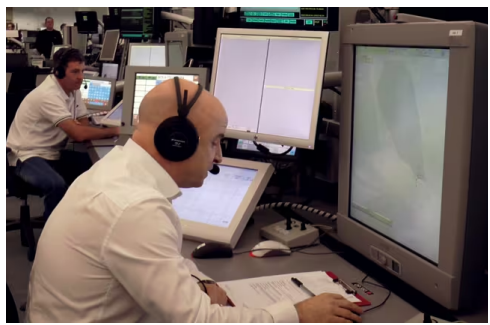

Figure 1: An ATCO controlling a sector. Directly to the front of the ATCO is the iTEC system, which includes a radar display and conflict detection functionality, with ancillary displays to the ATCO's left.

Members of a watch are co-located within the operations room, which affords collaboration, for example, on agreeing on a coordination. More informal kinds of collaboration are also important, however. One example is watch members' sharing of their experiences of using a particular tool. One ATCO remarked:

> "I think the controllers are pretty good at sharing, you know, tips on this or that… Anything that we've seen or that we like. I do all the time because you watch people, you see levels that they use or shortcuts that they use or don't use, you know, things that they've done… And you talk about why. And I think it's a constant evolution of your technique and your skills and how you use the tools, you know, things that they're really useful for."

In her study of air traffic controlling in the USA, Vaughan (2021) found that watches have distinctive cultures, which are reflected, *inter alia*, in differences in the ways that their members do their controlling work and how they use the tools available. Our study finds that is also the case of watches in UK air traffic control. An ATCO explained:

> "You do find that different watches use things in different ways, but that's always been the case. Different watches will interpret information or interpret how to use kit or how to use airspace differently so that's always happened, has always been slight differences in watches, but I think that the goal is still the same. The goal is still to make it the best version of controlling that you can do."

### 4.2 Trust and air traffic control work

The primary objective of an ATCO is to ensure the safety of the aircraft currently in their sector, that is to ensure that none of these aircraft are put at risk of colliding with one another. This objective is formalised by 'standards of separation': the minimum distance that must be maintained at all times between aircraft (vertical, lateral and longitudinal) or time interval. The distances and time intervals vary according to type of sector in which the aircraft is currently flying through.

An important aspect of being part of a watch is that members learn to trust each other's controlling skills. This is very important for safe and expeditious management of the aircraft for which the watch is collectively responsible. One ATCO explained:

> "It's an overall improved system when you have more trust in the people that you're working with and you're more familiar with them… Having the trust of your colleagues is a positive."

Safety is the primary objective of controlling, however, there are two other objectives that an ATCO will endeavour to meet given that safety is assured. These are orderliness and expeditiousness. The former relates to minimising the number of instructions (clearances) issued to aircraft (which has the benefit of reducing the ATCO's workload and the pilot's) and the latter relates to endeavouring to ensure aircraft pass through a sector as quickly as possible, so as to minimise the possibilities of delays.

ATCOs have a number of tools (see Figure 1 for an example) at their disposal to assist them in making sure that the aircraft they are controlling abide by these standards of separation. The tools are designed to satisfy the distinctive controlling demands of the different types of sectors. One example is providing warnings based upon predictions of potential violations of the standards of separation within a specific future time window. ATCOs learn from their experiences of using such tools that there are situations where it may be necessary to ignore a warning:

> "You can be overloaded by [the tool] because it will show a lot of potential conflicts with an aircraft that you wanted to climb… and at least half of the things that it would point out as a potential conflict were never conflicts. What you had to do was get good at weeding out the spurious conflict alerts to know what was real."

Another ATCO elaborated on how they learn from experience how to calibrate the trustworthiness of these tools:

> "We just know about its weakness… my experience of [the tool] 90% of what [the tool] does is spot on, it's just the odd thing where you think I'll be able to exercise a little bit of caution here. I've been caught out here before. Just watch this and not fully trust it."

## 5 DISCUSSION

The aim of this study is to contribute to understanding requirements for human-agent interaction in air traffic control and how meeting these requirements could afford ATCOs' trust in these agents, while maintaining their agency as decision-makers (Procter et al., 2023).

The findings reveal that though the tools ATCOs currently use to support their decision-making are not completely dependable and trustworthy in all circumstances, air traffic control work is nevertheless safe and dependable. What is important is that ATCOs are able to recognise and distinguish between dependable and undependable/relevant and irrelevant behaviours of the tools they use, that is to be familiar with their 'contours' of the trustworthiness, to have the situated awareness of where these contours lie and have knowledge of workarounds and skills to apply them in real-time so that they are able to continue to ensure the safety of the aircraft under their control. These knowledges and skills are acquired and honed through ATCOs' individual, everyday experiences of using these tools and through the sharing of experiences that goes on, for example, between watch colleagues.

Our findings suggest that it is not necessary for AI tools to be perfectly trustworthy and dependable for them to offer benefits to air traffic control. What is important is that ATCOs are able to recognise and navigate their trust contours and boundaries and thus know when – as well as how – to react at any particular moment.

## 6 CONCLUDING REMARKS

Grounded in a proven ethnographic methodology, with detailed field observations of subtle situated practices and interviews with practicing and trainee ATCOs, this paper documents how trust is practically achieved and maintained in the complex 'real-world' socio-technical system of air traffic control. Explicating the everyday, lived experience of air traffic controllers and instructors, and their subtle and nuanced notions of trust the article provides some valuable insights into trustworthy AI, particularly as agent-based AI systems are increasingly proposed for application in complex, safety-critical domains. Accordingly, the article introduces and illustrates the notion of 'trust contours' — the idea that trust is dynamic, locally calibrated and re-calibrated, and based on accumulated and shared experiences rather than simple fixed, global binary judgments. We suggest that this represents a meaningful conceptual advance in the discourse around explainable and trustworthy AI and provides useful insights for designing AI tools. We believe this kind of analysis is especially valuable for researchers, designers and practitioners developing AI for operational domains where the reliability of 'human-AI teaming' (Schmutz et al., 2024) is paramount.

Preparing ATCOs to work with agent-based tools will necessarily require training in their use and their behaviours, including their strengths and weaknesses. Developing explainable AI (xAI) techniques (Hughes et al., 2025) for agent-based tools so that ATCOs can associate these behaviours to the circumstances in which ATCOs encounter them will also be critically important. In other words, the behaviour of agents must be sufficiently transparent such that ATCOs can learn and dependably navigate their trust contours, identify and apply workarounds appropriate to these circumstances.

Our study reveals that ATCOs continue, as a natural element of their day-to-day work, to extend and refine their skills in assessing the trustworthiness of the tools they use and recognise and navigate their trust boundaries, gradients and contours. These informal processes of knowledge tuning and sharing are just as important as more formal processes, such as training. They are a vital element of a wider, socio-technical air traffic control system 'trust architecture', which ensures safe and dependable air traffic controlling in the face of the 'normal, natural troubles' that ATCOs deal with successfully every day.

Our ethnographic study of air traffic control work is continuing as part of a more detailed investigation of this trust architecture and the kinds of adaptations that the adoption of agent-based tools may require of it if their benefits are to be achieved fully, while maintaining a safe and dependable air traffic controlling service.

## ACKNOWLEDGMENTS

Project Bluebird is a Prosperity Partnership between NATS, the University of Exeter and the Alan Turing Institute for Data Science and AI, funded by the UK

Engineering and Physical Sciences Research Council (Grant Number EP/V056522/1).

**REFERENCES**

Anderson, R. (1994). Representations and requirements: the value of ethnography in system design. Human-computer interaction, 9(2), 151-182.

Bentley, R., Hughes, J. A., Randall, D., Rodden, T., Sawyer, P., Shapiro, D., & Sommerville, I. (1992). Ethnographically-informed systems design for air traffic control. In *Proceedings of ACM conference on Computer-supported cooperative work*, pp. 123-129.

Blackwell, A. F. (2021). Ethnographic artificial intelligence. Interdisciplinary Science Reviews, 46(1-2), pp.198-211.

Blazevic, R., Veledar, O., Stolz, M., & Macher, G., 2024. Toward safety-critical artificial intelligence (AI)-based embedded automotive systems. SAE International Journal of Connected and Automated Vehicles, 8 (12-08-01-0007).

Clarke, K., Hardstone, G., Rouncefield, M., & Sommerville, I. (Ed.). (2006). *Trust in technology: A socio-technical perspective* (Vol. 36). Springer Science & Business Media.

Garfinkel, H. (1967). *Studies in Ethnomethodology*, Englewood Cliffs, Prentice-Hall.

Dourish, P. (2006). Implications for design. In Proceedings of the SIGCHI conference on Human Factors in computing systems, pp. 541-550.

Hughes, J. A., Randall, D., & Shapiro, D. (1992, December). Faltering from ethnography to design. In *Proceedings of the 1992 ACM conference on Computer-supported cooperative work*, pp. 115-122.

Hughes, J. (2001). Of ethnography, ethnomethodology and workplace studies. *Ethnographic Studies*, 6, 7-16.

Hughes, L., Dwivedi, Y. K., Malik, T., Shawosh, M., Albashrawi, M. A., ... & Walton, P. (2025). AI Agents and Agentic Systems: A Multi-Expert Analysis. *Journal of Computer Information Systems*, 1-29.

Kaur, D., Uslu, S., Rittichier, K.J., & Durresi, A. (2022). Trustworthy artificial intelligence: a review. ACM computing surveys (CSUR), 55(2), pp. 1-38.

Luhmann, N. (2000). Familiarity, confidence, trust: Problems and alternatives. Trust: Making and breaking cooperative relations, 6(1), 94-107.

Marda, V., & Narayan, S. (2021). On the importance of ethnographic methods in AI research. Nature Machine Intelligence, 3(3), pp. 187-189

Procter, R., Tolmie, P., & Rouncefield, M. (2023). Holding AI to account: challenges for the delivery of trustworthy AI in healthcare. *ACM Transactions on Computer-Human Interaction*, *30*(2), 1-34.

Randall, D., Harper, R., & Rouncefield, M. (2007). Fieldwork for design: theory and practice. Springer Science & Business Media.

Randall, D., Rouncefield, M., & Tolmie, P. (2021). Ethnography, CSCW and ethnomethodology. Computer Supported Cooperative Work (CSCW), 30(2), pp.189-214.

Schmutz, J. B., Outland, N., Kerstan, S., Georganta, E., & Ulfert, A. S. (2024). AI-teaming: Redefining collaboration in the digital era. *Current Opinion in Psychology*, *58*, 101837.

Thornton, L., Knowles, B., & Blair, G. (2021). Fifty shades of grey: in praise of a nuanced approach towards trustworthy design. In Proceedings of the ACM Conference on Fairness, Accountability, and Transparency, pp. 64-76.

Vaughan, D. (2021). Dead reckoning: Air traffic control, system effects, and risk. University of Chicago Press.

Voß, A., Procter, R., Slack, R., Hartswood, M., & Rouncefield, M. (2006). Understanding and supporting dependability as ordinary action. *Trust in technology: A socio-technical perspective*, 195-216.

Watson, R. (2009). Constitutive practices and Garfinkel's notion of trust: Revisited. *Journal of Classical Sociology*, *9*(4), 475-499.